\documentclass[reprint,aps,twocolumn,showpacs,nofootinbib,superscriptaddress]{revtex4-2}

\usepackage{amsmath,amssymb,bm}
\usepackage[dvipdfmx]{graphicx}
\usepackage{hyperref}
\usepackage{yfonts}
\usepackage{color}
\usepackage{natbib}
\usepackage[normalem]{ulem}

\newcommand{\na}{{\bm \nabla}}

\newcommand{\xv}{{\bm x}}
\newcommand{\yv}{{\bm y}}

\newcommand{\Bv}{{\bm B}}

\newcommand{\e}{{\rm e}}
\renewcommand{\i}{{\rm i}}

\newcommand\xoutpars[1]{\let\helpcmd\xout\parhelp#1\par\relax\relax}
\newcommand\soutpars[1]{\let\helpcmd\sout\parhelp#1\par\relax\relax}
\long\def\parhelp#1\par#2\relax{%
  \helpcmd{#1}\ifx\relax#2\else\par\parhelp#2\relax\fi%
}

\begin{document}

\begin{flushright}KEK-TH-2512
\end{flushright}

\title{Novel transition dynamics of topological solitons}

\author{Kentaro Nishimura}
\affiliation{KEK Theory Center, Tsukuba 305-0801, Japan}
\affiliation{Research and Education Center for Natural Sciences, Keio University, 4-1-1 Hiyoshi, Yokohama, Kanagawa 223-8521, Japan}
\author{Noriyuki Sogabe}
\affiliation{Department of Physics, University of Illinois, Chicago, Illinois 60607, USA}
\affiliation{Quark Matter Research Center, Institute of Modern Physics, Chinese Academy of Sciences, Lanzhou, Gansu, 073000, China}

\begin{abstract}

Continuous phase transitions can be classified into ones characterized by local-order parameters and others that need additional topological constraints. The critical dynamics near the former transitions have been extensively studied, but the latter is less understood. We fill this gap in knowledge by studying the transition dynamics to a parity-breaking topological ground state called the chiral soliton lattice in quantum chromodynamics at finite temperature, baryon chemical potential, and an external magnetic field. We find a slowing down of the soliton's translational motion as the critical magnetic field approaches while the local relaxation rate to the solitonic state remains finite. The time required to converge to the stationary state strongly depends on the symmetry of the initial configuration, determining whether translational motion occurs during the dynamic process.

\end{abstract}
\maketitle

\section{Introduction}

\begin{figure}[t]
\centering
\includegraphics[bb=0 0 620 870, width=7cm]{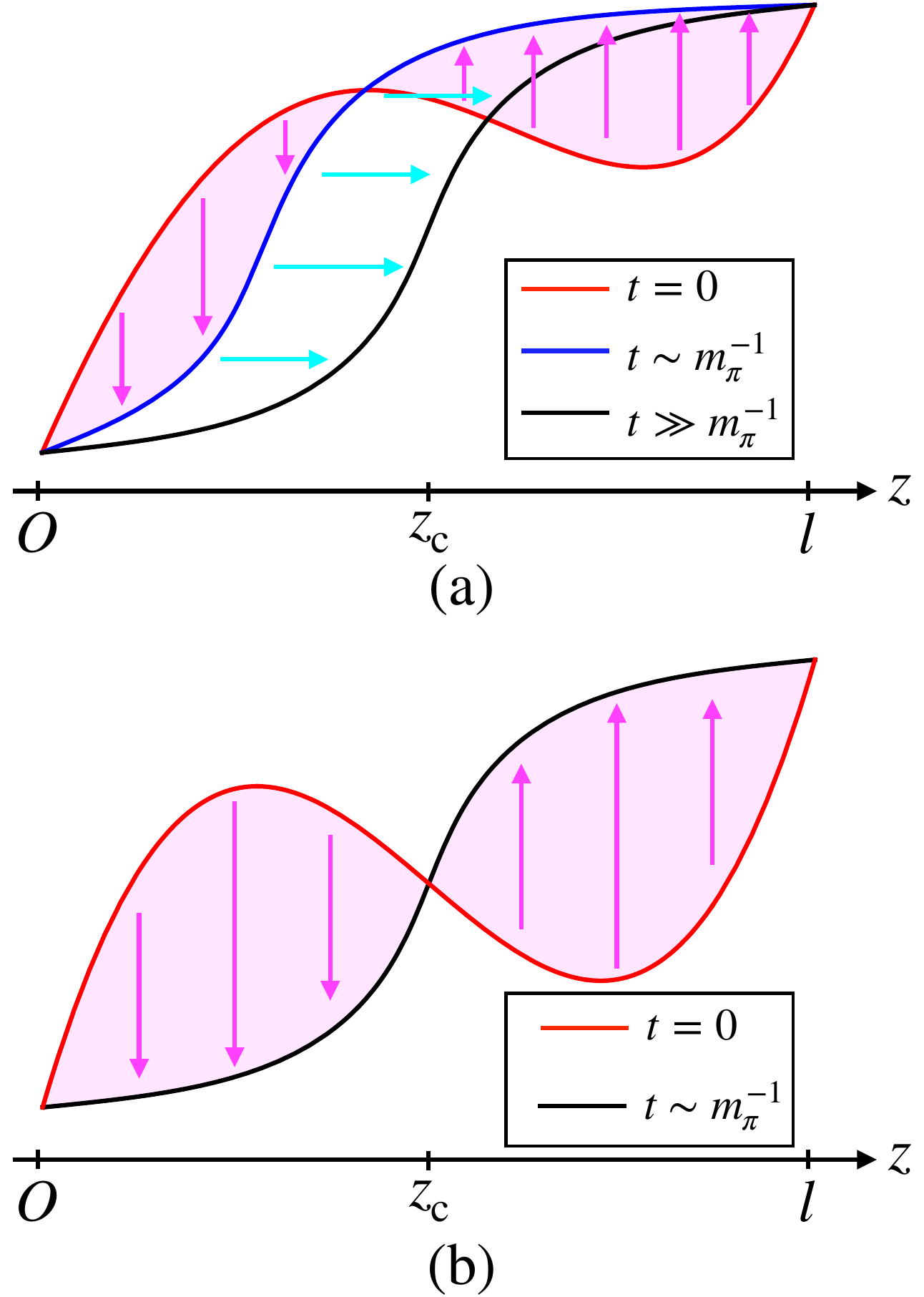}
\caption{Schematic illustration of the relaxation dynamics of the neutral pion field to its CSL state for a single lattice unit with the periodic length $l$. We consider two types of initial configurations in the modulation direction $z$: (a) a general (nonsymmetric) form and (b) a symmetric form (see main texts for details).}
\label{fig:illustration}
\end{figure}

The most traditional way of classifying phase transitions is by distinguishing between the discontinuous (first-order) and continuous (second-order) types. A further classification has been proposed based on topology \cite{deGennes1975} for the latter. Standard ones are topologically trivial and can be described by local-order parameters, such as the magnetization of a ferromagnet. However, there are also continuous transitions in which the characteristic order parameters cannot be defined similarly but with additional topological constraints, such as the transition between a type-II superconductor's Meissner and mixed states, which involves a change in the topological number of Abrikosov vortices \cite{abrikosov1957magnetic} (see also Ref.~\cite{SCvol2}). Trivial transitions are characterized by scaling power laws, while nontrivial ones have a ubiquitous logarithmic behavior either below or above the transition point, e.g., in chiral magnets \cite{Dzyaloshinsky}, cholesteric liquid crystals \cite{de1993physics}, etc.

Topologically trivial second-order phase transition dynamics have been extensively studied \cite{RevModPhys.49.435}, but relatively little is known about the nontrivial transitions. One general phenomenon of trivial transitions is the critical slowing down of hydrodynamic modes coupled with the order parameter or the order parameter itself (see \cite{chaikin_lubensky_1995, HOHENBERG20151}), such as the vanishing diffusion coefficients of a binary fluid near the liquid-gas critical point. Is any singular dynamical behavior exhibited by topological objects, including domain walls and vortices, near the transition?

This paper presents the first study of the relaxation dynamics of reaching the chiral soliton lattice (CSL), a parallel stack of domain walls with a topological charge. Similar solitonic structure has also been observed in various condensed matter systems \cite{Dzyaloshinsky,de1993physics} and in quantum chromodynamics (QCD) under an external magnetic field \cite{Brauner:2016pko,Son:2007ny,Eto:2012qd,Brauner:2017mui,Brauner:2021sci,Brauner_2023}, rotation \cite{Huang:2017pqe,Nishimura:2020odq,Eto:2021gyy}, time-periodic circularly polarized laser \cite{Yamada:2021jhy}, and in QCD-like theories \cite{Brauner:2019aid,Brauner:2019rjg}. We examine QCD under specific conditions: finite temperature $T$, finite baryon chemical potential $\mu_{\rm B}$, and an external magnetic field $\Bv$. This exploration is motivated by potential manifestations of inhomogeneous states within magnetar cores and during noncentral heavy-ion collisions (see also a recent proposal in Ref.~\cite{Fukushima:2023tpv}). Given the initially large magnetic fields in these systems and their subsequent decay, a potential exists for realizing the CSL transiently during dynamic processes. Concurrently, there is a need to establish a dynamic framework incorporating the CSL states known in the static theory. However, to our knowledge, such a comprehensive theory has yet to be developed.


Figure~\ref{fig:illustration} (a) schematically illustrates the relaxation dynamics of the neutral pion field to its CSL state with two characteristic features. The first is the local relaxation dynamics toward the domain wall, represented by the magenta region and arrows transitioning from the red to blue configurations. The second feature is the translational motion of the domain wall, often termed motion in the moduli space, which is depicted by cyan arrows moving from the blue to the black configurations. In this illustration, the red and black configurations signify a general initial state and a stationary CSL unit, respectively. An intermediate state is described by the blue configuration, a translating domain wall positioned away from the stationary state. This transient state has a slightly higher energy than the stationary state, prompting the system to evolve toward the stationary configuration.

Our analysis shows a pronounced timescale separation between these two characteristic ``modes'' as the transition to the CSL approaches. The local relaxation time remains finite as the transition point, $t\sim m_\pi^{-1}$ at the order of the inverse pion mass, whereas the typical completion time for translational motion increases significantly, $t\gg m_\pi^{-1}$. These behaviors indicate that the presence or absence of translational motion in the dynamical process considerably impacts the duration required to achieve a stationary state.

In Figs.~\ref{fig:illustration} (a) and \ref{fig:illustration}(b), we juxtapose the dynamics of two initial configurations, $\phi_{\rm ini}(z)$: (a) represents a general (nonsymmetric) form, while (b) depicts an odd function at the center of the domain wall in the stationary state $z_{\rm c}$, i.e., $\phi_{\rm ini}(z-z_{\rm c}) - \phi_{\rm ini}(z_{\rm c})=-\phi_{\rm ini}(z_{\rm c}-z) + \phi_{\rm ini}(z_{\rm c})$. Owing to this symmetry, the relaxation process in (b) avoids the translational motion of the kink. This is because it already aligns with the stationary state's position from the outset, as evidenced by ${\rm d}^2 \phi_{\rm ini}/{\rm d} z^2 =0$ at $z=z_c$. We will further illustrate and quantify these behavioral differences using examples in Figs.~\ref{fig:evo} and \ref{fig:R}. Our findings reveal a unique facet of second-order phase transition dynamics driven by topological solitons.

This paper is organized as follows: In Sec.~\ref{sec:setup}, we present our setup, incorporating both static and dynamic formulations. In Sec.~\ref{sec:statics}, we analyze the stationary solutions and their alterations due to dissipation. In Sec.~\ref{sec:dynamics}, we present our numerical results on the time evolution of the system, highlighting the rapid and slow processes leading to the CSL state. Section \ref{sec:dis} discusses the physical origins of the time scale separation and its implications in tabletop experiments of chiral magnets. We conclude in Sec.~\ref{sec:conc}. Appendix \ref{sec:ana} is dedicated to establishing a connection between the terminologies of QCD and helimagnets within the CSL context.

\section{Formulation}
\label{sec:formulation}
\subsection{Setup}
\label{sec:setup}

We consider an effective theory of two-flavor QCD at finite $T$, $\mu_{\rm B}$, and $\Bv$ such that CSL appears as the ground state \cite{Brauner:2016pko, Brauner:2021sci}. Our theory takes into account low-energy degrees of freedom, such as the Nambu-Goldstone (NG) modes and conserved charge densities that respect the chiral symmetry, i.e., the neutral pion $\phi$ with $\bar q q \sim \e^{2 \i \phi t^3}$ and the axial isospin charge density $\rho = \bar q \gamma^0 \gamma^5 t^3 q$, where $q$ is the quark field, and $t^a$ is the ${\rm SU}(2)$ generator with ${\rm tr} (t^a t^b) = \delta^{ab}/2$. Note here that $\phi$ is the (pseudo) NG mode associated with the spontaneous symmetry breaking of the axial $t^3$ rotation of the chiral symmetry, ${\rm SU}(2)_{\rm L} \times {\rm SU}(2)_{\rm R}$, denoted by ${\rm U}_{\rm L-R}(1)^{t^3}$, which is an approximate symmetry in the presence of the quark mass. $\rho$ is the conserved charge density for the same approximate symmetry, ${\rm U}_{\rm L-R}(1)^{t^3}$. Near the transition between the QCD vacuum and the CSL state, we can discard the charged pions $\pi^\pm$ with an additional mass due to $\Bv\neq 0$ and the other conserved charge densities (see Refs.~\cite{Brauner:2016pko, Eto:2023lyo} for interesting phenomena among $\pi^\pm$). However, to describe the dynamics near the transition to the nuclear matter at $\mu_{\rm B}$ comparable with the nuclear mass, we may need to include the energy and momentum densities (see Ref.~\cite{Son:1999pa}). We set $e=1$.

\subsection{Statics}

The Hamiltonian density of the system, $\mathcal H = \mathcal H _\phi + \mathcal H _\rho $ consists of the pion and charge sectors:
\begin{subequations}
\label{eq:ham_a=3}
\begin{align}
\label{eq:ham_a=3-1}
\mathcal H _\phi &= \frac{f_\pi^2}{2} (\na \phi)^2
-f_\pi^2 m_{\pi}^2 \cos \phi -\frac{\mu_{\rm B}}{4 \pi ^2}\Bv \cdot \na \phi \,, \\
\label{eq:ham_a=3-2}
\mathcal H _\rho &= \frac{1}{2 \chi} \rho^2 \,.
\end{align}
\end{subequations}
Here, the pion decay constant $f_\pi$ characterizes the stiffness of the chiral order, $m_{\pi}$ is the pion screening mass, and $\chi$ is the axial isospin charge susceptibility. The first two terms on $\mathcal H _\phi$ are the kinetic and mass terms, which can be derived based on the chiral symmetry and its explicit breaking by a small quark mass; the third one is a topological term due to the anomalous coupling of the neutral pion and $\Bv$ in the presence of $\mu_{\rm B}$ \cite{Son:2007ny,Son:2004tq}. The transition we are interested in occurs through the competition between the kinetic and topological terms, which disfavor and favor inhomogeneity, respectively. We have neglected any nonlinear and spatially varying terms in the density term $\mathcal H _\rho$.

\subsection{Dynamics}

We employ the Poisson bracket method to derive the hydrodynamic equations that follow from the system's symmetries (for general construction, see Refs~\cite{chaikin_lubensky_1995, DZYALOSHINSKII198067})
\begin{subequations}
\label{eq:EOM-gen}
\begin{align}
\partial_t \phi(\xv) &= \int {\rm d} \yv \left[ \phi(\xv), \rho(\yv) \right] \frac{\delta H}{\delta \rho(\yv)} - \kappa \frac{\delta H}{\delta \phi(\xv)} \,, \\ 
\label{eq:EOM-gen2}
\partial_t \rho(\xv) &= \int {\rm d} \yv \left[ \rho(\xv), \phi(\yv) \right] \frac{\delta H}{\delta \phi(\yv)} + \lambda \na^2 \frac{\delta H}{\delta \rho(\xv)} \,,
\end{align}
\end{subequations}
where $H=\int {\rm d}\xv\, \mathcal H$, and $\kappa$ and $\lambda$ are relaxation rate and axial isospin conductivity. In each of (\ref{eq:EOM-gen}), the first and the second terms describe the macroscopic Hamiltonian dynamics and the thermal dissipation, which preserves and breaks the time-reversal symmetry, respectively. We posturate a Poisson bracket such that $\phi$ and $\rho$ are canonical conjugates, $\left[ \phi(\xv),\rho(\yv) \right] = \delta (\xv-\yv)$, consistent with the ${\rm U}(1)_{t^3}$ symmetry \cite{Son:2000ht, weinberg_1996}: $\mathcal L=\mathcal L(\partial_t \phi + \mu_{\rm A},\na \phi)$, hence $\rho \equiv \delta \mathcal L/\delta \mu_{\rm A} = \delta \mathcal L / \delta (\partial_t \phi)$. The coefficients of the functional derivative in the dissipative terms are determined by a derivative expansion. The leading contribution of \eqref{eq:EOM-gen2} is $\mathcal O(\na^2)$ due to the approximate axial isospin conservation. For simplicity, we neglect noise terms and the anisotropy in the conductivity.

By performing functional derivatives, we arrive at the following expression:
\begin{subequations}
\label{eq:eom}
\begin{align}
\label{eq:eom_phi}
\partial_t \phi &= \frac{\rho}{\chi} + f_\pi^2 \kappa\left(-m_{\pi}^2\sin \phi +\na^2 \phi \right)\,,\\
\label{eq:eom_rhoA}
\partial_t \rho&=f_\pi^2 \na^2 \phi - m_\pi^2f_\pi^2 \sin \phi +\frac{\lambda}{\chi} \na^2 \rho \,,
\end{align}
\end{subequations}
which represents a nonlinear and dissipative extension of previous studies \cite{Son:1999pa, Son:2002ci}.  It is worth noting that both features are crucial for the system to converge to the correct stationary state in the presence of an infinite potential of the form $\propto \cos \phi$ rather than a mass term $\propto \phi^2$.

We set $\Bv = B {\bm e}_z$ and use units of $m_\pi^{-1}$ and $M_\pi^{-1}$ for space and time coordinates, respectively. Here, $M_\pi = m_\pi c$ represents the pion pole mass, where $c^2 =f_\pi^2/\chi$ denotes the pion velocity in the chiral limit \cite{Son:2002ci}. Note that the system is homogeneous in the $xy$ plane as $\partial_{x,y}$ only enters the first term of $\mathcal H_\phi$. We first eliminate $\rho$ and neglecting $\mathcal O(\partial_z^4)$ from the original equations \eqref{eq:eom}, we obtain one of the main equations of this paper:
\begin{align}
\label{eq:EOM}
& \square \phi + \sin \phi  + \gamma \partial_t \sin \phi  - \alpha \partial_z^2 \partial_t \phi - \beta \partial_z^2 \sin \phi =0 \,,
\end{align}
where $\square \equiv \partial_t^2 -\partial_z^2$, $\gamma = m_\pi f_\pi^2 \kappa /c$, $\alpha = \gamma + m_\pi \lambda/(c\chi) $, and $\beta = \kappa \lambda m_\pi^2$. We can track the dynamics of $\rho$ by substituting $\phi$ back into the following dimensionless form of Eq.~(\ref{eq:eom_phi}):
\begin{align}
\label{eq:EOM-charge}
\frac{\rho}{m_{\pi} \chi} = \partial_t \phi + \gamma \left(\sin \phi - \nabla^2 \phi \right) \,.
\end{align}

\section{Chiral soliton lattice and dissipation}
\label{sec:statics}

\subsection{Stationary equation}

The stationary solution of the time-evolution equation (\ref{eq:EOM}), $\bar \phi(z)$ contains the spatial profile of the CSL given by 
\begin{align}
\label{eq:ss-eq}
\partial_z^2 \bar \phi = \sin \bar \phi - \beta \partial_z^2 \sin \bar \phi \,.
\end{align}

\subsection{Dissipationless solution}
Let us first examine the dissipationless limit, $\beta=0$ (see \cite{Brauner:2016pko} for details). This equation has the same mathematical form as the simple pendulum, and we can solve it analytically using Jacobi's elliptic function (a dimensionless form):
\begin{gather}
\label{eq:ss_sol_beta_0}
\bar{\phi}(z, k) = 2 \textrm{am}\left(\frac{z}{k},k \right) + \pi \,,
\end{gather}
where $k\, (0\leq k \leq 1)$ is the elliptic modulus, a free parameter at this stage. Note that this solution (\ref{eq:ss_sol_beta_0}) has periodic minima at $\bar{\phi}((2m+1)kK(k),k) = 2\pi m$, with $m$ being an integer and the periodic length is given by $\ell=2kK(k)$. Here, $K(k)$ and $E(k)$ (for later purposes) are the complete elliptic functions of the first and second kinds, respectively.

We now determine the optimal $k$ for given $B$ with fixed $\mu_{\rm B}$ by minimizing the total energy in $z$ direction per unit area in the $xy$ plane, $\int_0^l {\rm d} z\mathcal H _{\phi = \bar \phi}/l$ with the solution (\ref{eq:ss_sol_beta_0}). After a straightforward calculation, we find the minimization condition reduces to $E(k)/k= B/B_{\rm CSL}$ with $B_{\textrm{CSL}}=16\pi m_{\pi}f_{\pi}^2/\mu_{\rm B}$. The solution $k(B)$ exists if and only if $E(k)/k\ge 1$ and the equality holds in the critical magnetic field, $B_{\rm CSL}$, which separates the QCD vacuum $(B<B_{\rm CSL})$ and the CSL $(B>B_{\rm CSL})$.
By substituting $k=k(B)$ into $\ell$,
we numerically determine the $B$ dependence of the periodic length, $\ell(B)$.

Near $B_{\rm CSL}$ from above, an asymptotic form of the complete elliptic functions ($k'=\sqrt {1-k^2}$) can be used  \cite{whittaker_watson_1996}:
\begin{subequations}
\begin{align}
K(k)& \simeq \log \frac{4}{k'} + \frac{{k'}^2}{4}\left( \log \frac{4}{k'}-1\right)\,, \\
E(k)& \simeq 1 + \frac{{k'}^2}{2}\left( \log \frac{4}{k'}-\frac{1}{2}\right)\,.
\end{align}
\end{subequations}
The energy minimization condition reduces to $k \simeq B_{\rm CSL}/B$ and thus $l(B) \simeq - \log (\Delta B/B)$ with $\Delta B \equiv B-B_{\rm CSL}$. The leading logarithmic behavior can also be derived from the balance between the repulsion energy of the kinks separated by $l$ \cite{PERRING1962550} (see also Ref.~\cite{Manton:2004tk}) and the binding energy gain from the topological term measured from $B=B_{\rm CSL}$: 
\begin{align}
\label{eq:l-B}
\e^{-lm_\pi} \sim \frac{\mu_{\rm B} \Delta B }{4\pi^2} \int_0^l{\rm d}z \partial_z \phi = \frac{\mu_{\rm B}\Delta B}{2\pi} \,, 
\end{align}
where we have restored the unit length $m_{\pi}^{-1}$. The bulk quantities reflect the logarithmic behavior of $l(B)$, e.g., the baryon number per unit area in the $xy$ plane,
\begin{align}
\label{eq:tot-N}
(N_{\rm B})_{\rm tot} = \frac{N_{\rm B}}{l} \sim -\frac{1}{\log (\Delta B/B)}\,,
\end{align}
where $N_{\rm B}=-\int_0^l {\rm d}z \partial \mathcal H _{\phi = \bar \phi}/(\partial \mu_{\rm B}) = B/(2\pi) $ is the baryon number per unit CSL \cite{Brauner:2016pko,Son:2007ny}. By integrating $\partial \mathcal H _{\phi = \bar \phi}/(\partial B)$ instead of $\partial \mathcal H _{\phi = \bar \phi}/(\partial \mu_{\rm B})$, the same behavior of the magnetization can be checked. 
Note that, even when taking into account the quantum fluctuations of the pions, the total number density [as given in Eq.~(\ref{eq:tot-N})] and magnetization near the critical magnetic field maintain characteristics of the continuous transition with a singular derivative. This has been recently demonstrated in Ref.~\cite{Brauner_2023}
.


\subsection{Dissipative solution}

\begin{figure}[t]
\centering
\includegraphics[bb=0 0 670 390, width=8cm]{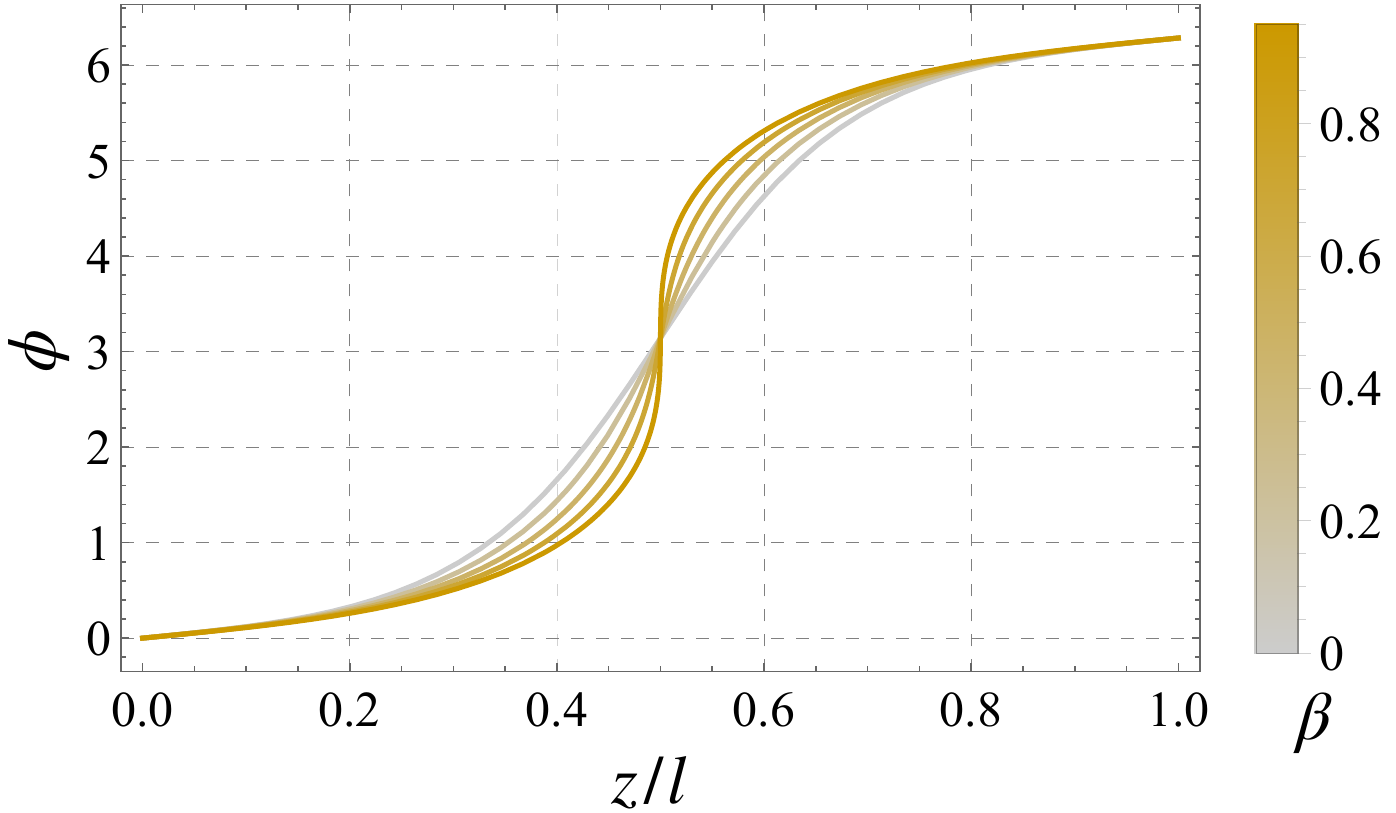}
\caption{Static configurations of the neutral pion field $\bar \phi$ described by Eqs.~(\ref{eq:ss-eq}). Variations for different strengths of dissipation $\beta$ are shown. A unit of CSL is depicted, with the spatial coordinate $z$ normalized by the period of the CSL, $l$.}
\label{fig:ss}
\end{figure}

Now let us consider the dissipative case $\beta \neq 0$, applied to the stationary solution (\ref{eq:ss-eq}). Figure~\ref{fig:ss} (a) shows the $\beta$ dependence of a unit CSL configuration of the $\bar \phi$ field with the boundary condition, $\bar \phi (0)=0$ and $\bar \phi (l)=2 \pi$ for an arbitrary $B>B_{\rm CSL}$. As $\beta$ increases, the domain wall at the center sharpens, corresponding to a reduction in the width of the kink in the $\partial_z \bar \phi$ configuration. Interestingly, nonlinear effects arise in the dissipation of the CSL, $\propto \partial_z^2 \sin \bar \phi \neq \partial_z^2 \bar \phi$. If one neglects the nonlinearity from the right-hand sides of the dynamical equation (\ref{eq:EOM}), $\sin \phi \approx \phi$ leads to a different convergence of the pion field. For the convergence to occur, $\beta$ must be less than 1; otherwise, the solution becomes unstable at $\beta\geq 1$.

\section{Relaxational dynamics to the chiral soliton lattice}
\label{sec:dynamics}

\subsection{Numerical setup}
We numerically study the dynamical relaxation processes toward the stationary state discussed so far by solving Eq.~(\ref{eq:EOM}). We consider a finite length in the $z$ direction ($0\leq z \leq L$) and vary $B$ such that $l(B)=L/n$ discontinuously \cite{comment:cont:B}, where $n=1,2,\dots$ corresponds to the number of kinks in the system. We discuss a possible problem in the infinite system later. We impose each of the initial-boundary conditions as follows:
\begin{subequations}
\label{eq:ini_con}
\begin{align}
\text{(I)} & \quad 
\phi(0,z)=0\,, \quad \phi(t,L)=2 n \pi \theta (t-t_0) \,,\\ 
\label{eq:ini_con-II}
\text{(II)} & \quad \phi(0,z)=2n \pi z/L\,, \quad \phi(t,L)=2n\pi\,,
\end{align}
\end{subequations}
which have a common ground state at $t\rightarrow \infty$. Note that minimizing the total energy determines $n$ at fixed $L$ and $B$. In this sense, the information of the magnetic field is implemented as the boundary condition. Case (I) describes a quench from the vacuum to the CSL state at $t=t_0$, while case (II) describes the relaxation process starting with a symmetric inhomogeneous configuration that is not in the ground state. Here, we have selected representative initial configurations: (I) with and (II) without translational motion in the subsequent dynamic process. We emphasize here that instead of the liner configuration adopted in (\ref{eq:ini_con-II}), one can choose any initial configurations that exhibit odd symmetry about the center at $z_c=(i-1/2)L/n$ for an arbitrary $i$th cell $(i=1,\dots,n)$, defined as $(i-1)L/n<z<iL/n$. Owing to this symmetry, the kink emerges at each center and does not move throughout the entire dynamic process. Throughout the simulations below, we set the parameters $(\alpha,\,\beta,\,\gamma) = (2.2,\, 0.4,\,0.2)$. 

\subsection{Time evolution} 

\begin{figure}[t]
\centering
\includegraphics[bb=0 0 900 850, width=8.5cm]{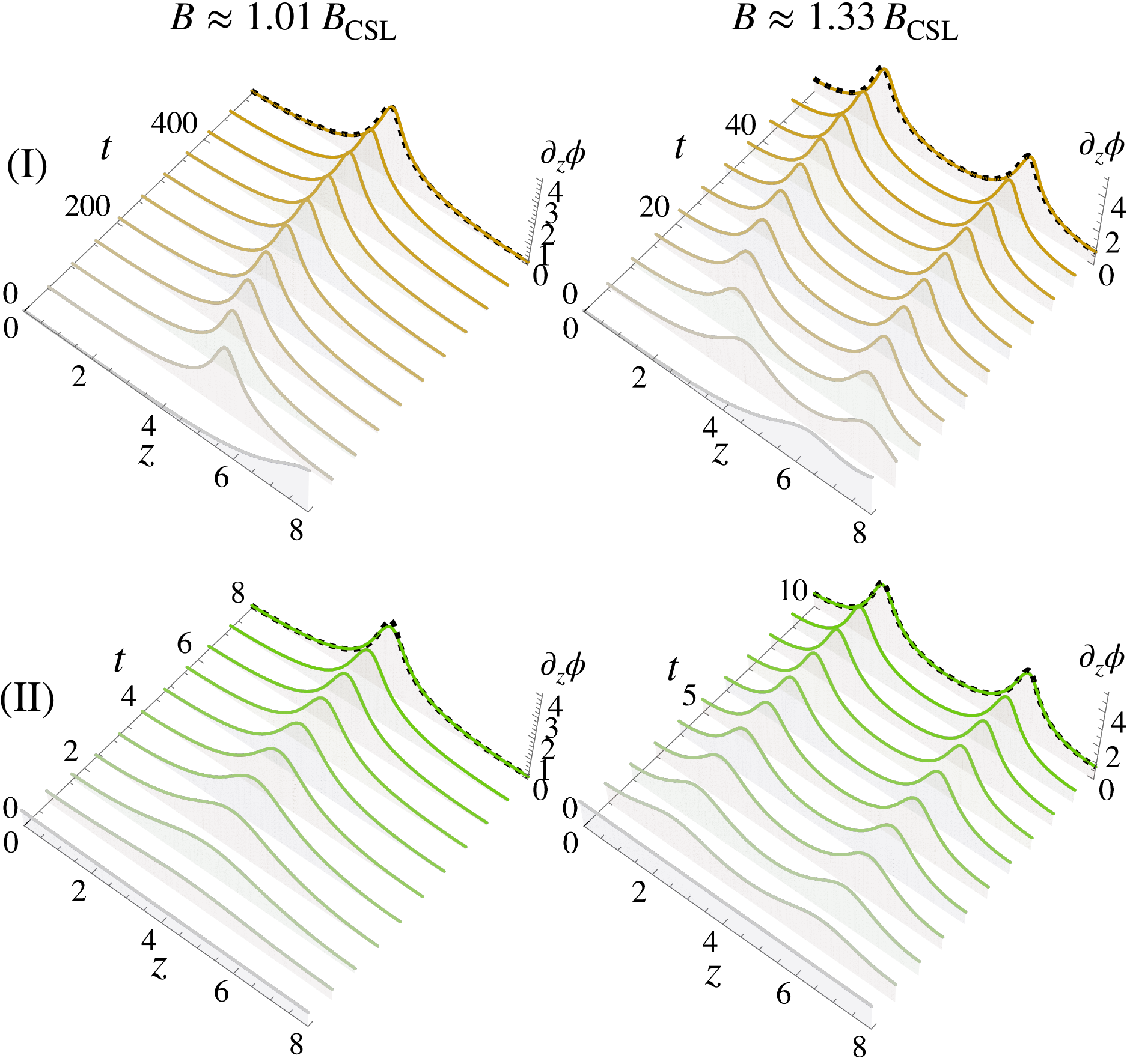}
\caption{
Time evolution of $\partial_z \phi$ for different combinations of initial-boundary conditions (I) or (II) and external magnetic field, $B\approx 1.01\,B_{\rm CSL}$ or $B \approx 1.33\,B_{\rm CSL}$. The black dashed curves represent the stationary-state solution (\ref{eq:ss-eq}), corresponding to Fig.~\ref{fig:ss} at $\beta =0.4$. The plotted data are limited to a time where the deviation from the stationary solution becomes less than $1\%$ of the initial deviation.
}
\label{fig:evo}
\end{figure}

Figure~\ref{fig:evo} shows a typical time evolution for each of the initial-boundary conditions (\ref{eq:ini_con}) with two values of $B=l^{-1}(L/n)$ $(n=1,2)$ for $L=8.0$, which is sufficiently larger than the typical size of the soliton ($\sim 1.0$ in the unit of $m_\pi^{-1}$) so that we can neglect boundary effects. The yellow curves correspond to case (I), where the solitonic object (for each $n=1$ and $n=2$) is initially created at the edge $z=L$ and moves rigidly to the ground-state position as time progresses. The green curves correspond to case (II), where the initial uniform configuration of $\partial_z \phi$ approaches the stationary configuration without any translational motion in the moduli space and only by the local relaxation. As the time intervals are different orders of magnitudes, case (I) at $B\simeq B_{\rm CSL}$ (top left) takes significantly longer compared to the others: case (I) relatively far from $B_{\rm CSL}$ (top right) and case (II) for any $B$ (bottoms).

\subsection{Relaxation rate}

\begin{figure}[t]
\centering
\includegraphics[bb=0 0 580 380, width=8cm]{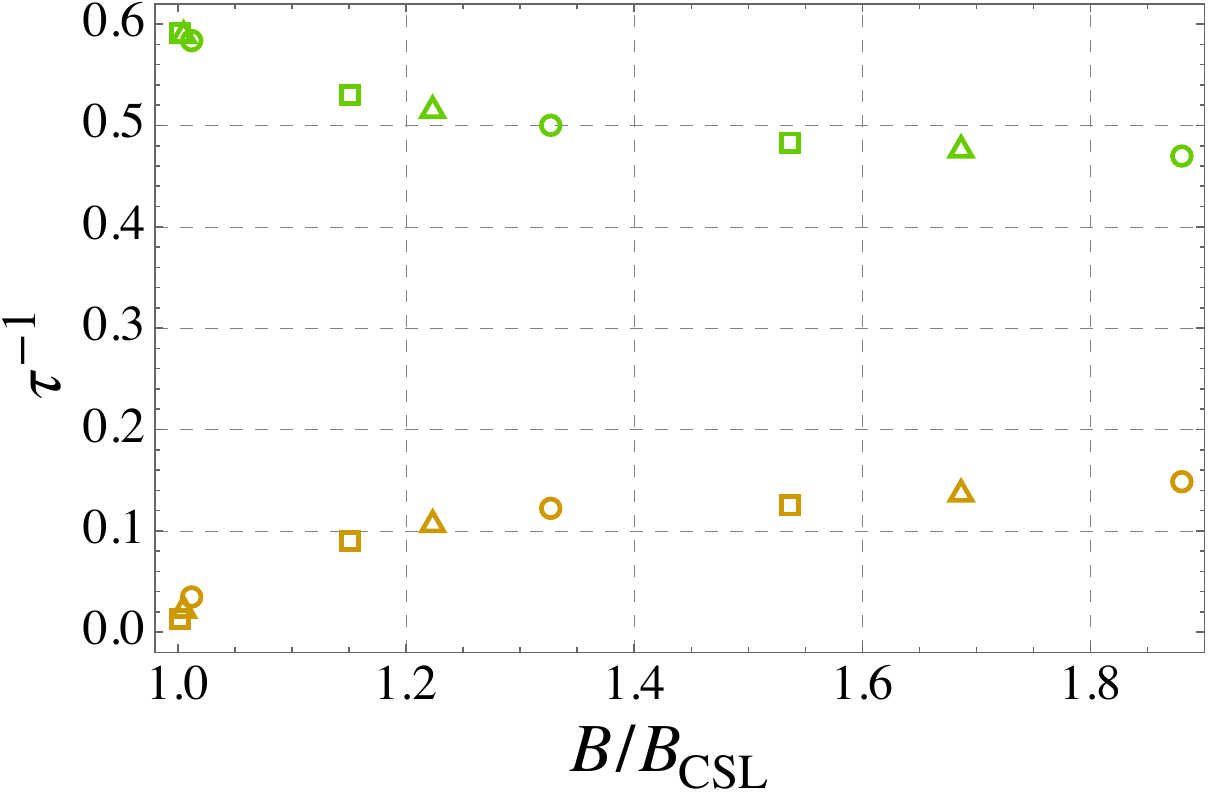}
\caption{
Dependence of the characteristic rate $\tau^{-1}$ on the magnetic field for each initial-boundary condition, (I) with and (II) without translational motion in the dynamical process, represented in yellow and green curves, respectively. The symbols (circle, triangle, and square) represent different system sizes, $L=8,9$, and $10$, respectively. As the applied magnetic field $B$ approaches to the critical value $B_{\rm CSL}$, the number of kinks in the system ($n=1,2,3$) decreases, leading to a slower (yellow) and faster (green) approach to the stationary state.}
\label{fig:R}
\end{figure}

To quantify the soliton's slow motion in Fig.~\ref{fig:evo}, we define the displacement from the stationary state $\bar \phi(z)$ given by Eq.~(\ref{eq:ss-eq}): $\Delta \Phi (t) \equiv 
\int_0^L {\rm d} z \left|\phi(t,z)-\bar \phi(z)\right|/Ln\pi$, 
where the normalization is chosen so that $\Delta \Phi (0)=1$. We obtain the characteristic time $\tau$ such that the deviation decays ``expositionary,'' $\Delta \Phi (\tau)={\rm e}^{-1}$, and evaluate how fast the system approaches the stationary state using the characteristic rate $\tau^{-1}$. Figure~\ref{fig:R} shows the dependence of this rate on the magnetic field for each initial-boundary condition \eqref{eq:ini_con} and different system size $L$. For case (I), the rate tends to $0$ as $B/B_{\rm CSL}\rightarrow 1$, whereas it remains finite for case (II). The first two circles ($L=8$) for each yellow and green plot correspond to the demonstration in Fig.~\ref{fig:evo}, for $B=1.01\, B_{\rm CSL}$ and $B=1.33\, B_{\rm CSL}$, respectively. Note that the behavior as a function of $B/B_{\rm CSL}$ is independent of $L$, although it will generally break down in a large $B$ region, where the spatial separation of the soliton is insufficient, $l(B)\sim 1$.

\section{Discussion}
\label{sec:dis}

As depicted in Fig.~\ref{fig:R} (see also Fig.~\ref{fig:evo} for a demonstration), we observe a slowing down of the translational motion of the soliton as the external magnetic field $B$ approaches the transition value $B_{\rm CSL}$. Interestingly, the local relaxation rate to the domain wall remains finite as $B \rightarrow B_{\rm CSL}$. This is due to the singularity at $B=B_{\rm CSL}$ characterized by the divergence of the CSL separation length, $l(B)\rightarrow \infty$, which only affects the solitonic motion. Specifically, the yellow plots reach exactly zero at $B=B_{\rm CSL}$, consistent with the fact that the kink remains stationary in the exact infinite volume, where the translational invariance is restored. We attribute the finite relaxation process to the excited-state dynamics of the soliton, whose typical frequency $\omega \simeq m_{\pi}$ remains gapped even at $B=B_{\rm CSL}$ \cite{sutherland1973some,whittaker_watson_1996,landau2013quantum} (see also Ref.~\cite{KISHINE20151} for the review).

We can extend this analysis to an infinite system perturbed locally by a single CSL unit. We assume the initial deformation does not affect the other configurations, which are kept as ground states. This assumption is justified near $B_{\rm CSL}$ since $l(B)$ is large and the nearest neighbors are sufficiently separated. We can consider similar initial configurations as we discussed previously [see Eq.~(\ref{eq:ini_con}) with $n=1$] and vary $B$ continuously. The characteristic time to reach a stationary state will have a magnetic field dependence similar to that shown in Fig.~\ref{fig:R}, where cases (I) and (II) correspond to the dynamics resulting from an initial dislocation and deformation of the kink, respectively. It would be interesting to test these predictions, for instance, in chiral magnets \cite{PhysRevLett.108.107202} (see Appendix.~\ref{sec:ana} for an analogy between the CSL in QCD and chiral magnets).

\section{Conclusion.}
\label{sec:conc}

We have presented a dissipative hydrodynamic framework with a small quark mass for QCD at finite temperature, baryon chemical potential, and the external magnetic field [see Eq.~(\ref{eq:eom})]. Our effective theory incorporates a nonlinear representation of the neutral pion field and successfully has a stational solution associated with a topological ground state, known as the CSL state. This would be the first step to describe the dynamics of the CSL in noncentral heavy-ion collisions and the cores of magnetars. It would be interesting to apply our model to the quench dynamics of the transient CSL state sustained only for a short time by a sufficiently strong magnetic field.

Our numerical simulations reveal a slowing down of the soliton's translational motion in the moduli space near the transition point between topologically trivial/nontrivial states while the local relaxation to the topological soliton keeps finite in the same region (see Figs.~\ref{fig:evo} and \ref{fig:R}). We propose that this dynamical characteristic is not exclusive to the CSL but may represent a general feature of second-order transitions involving a change in topological number. We expect analogous slow and fast dynamics by considering the relaxation process to a topological object and controlling the motion in the moduli space by imposing the appropriate symmetry on the initial condition. This feature contrasts with the standard trivial second-order phase transitions, where both collective motions and local dissipation show simultaneous critical slowing down due to the large correlation length of the order parameter.

\begin{acknowledgements}
N. S. thanks Misha Stephanov, Ho-Ung Yee, Maneesha Sushama Pradeep, and Masaru Hongo for the useful discussion and Yuya Tanizaki, Igor Shovkovy, and Haruki Watanabe, for fruitful comments and questions. We thank Yoshimasa Hidaka for the comments and questions on the manuscript. We thank Naoki Yamamoto for the useful discussion. K. N. is supported by JSPS KAKENHI, Grant-in-Aid for Scientific Research No. (B) 21H01084. N. S. acknowledges the support from the Strategic Priority Research Program of Chinese Academy of Sciences, Grant No. XDB34000000. This work is supported by the U.S. Department of Energy, Office of Science, Office of Nuclear
Physics Award No. DE-FG0201ER41195.
\end{acknowledgements}

\appendix

\section{Analogy between CSL in QCD and chiral magnets}
\label{sec:ana}

The CSL is also manifested in a certain class of helimagnets, which are stabilized by the relativistic spin-orbit coupling known as the Dzyaloshinskii-Moriya (DM) interaction. In this system, spins align within an easy plane (the $xy$ plane) and undergo periodic rotation along the normal $z$ direction. This rotation is generated by the DM interaction, which promotes the orthogonal alignment of adjacent spins. Within the easy plane, rotational symmetry is spontaneously broken. The corresponding NG mode is the magnon, characterized by the spin angle in the easy plane. The conserved charge associated with this rotational symmetry is identified as the longitudinal magnetization density along the $z$ axis. The total spin in the $z$ direction is a symmetry generator within this easy plane. For a pedagogical review of general easy plane magnet dynamics, readers may refer to Sec.~3.3.4 of Ref. \cite{täuber_2014}.

Utilizing the correspondence between the symmetry groups ${\rm U}(1) \simeq {\rm O}(2)$, we can map the effective Hamiltonian of the CSL in QCD to that of chiral magnets as follows: First, the Hamiltonian of the magnon along the $z$ direction exhibits the same form as that of the pion, as denoted by Eq.~(\ref{eq:ham_a=3-1}) [see also Eq.~(45) in Ref.~\cite{KISHINE20151}]. In this mapping, the pion mass term is analogous to an external symmetry-breaking term, represented by the magnetic field within the easy plane to say in the $x$ direction. The anomaly and DM interaction terms exhibit equivalent mathematical behavior despite having different physical origins. Second, the Hamiltonian of the longitudinal spin is equivalent to that described by the axial isospin charge, as represented by Eq.~(\ref{eq:ham_a=3-2}). The corresponding Hamiltonian in chiral magnets is identified as the anisotropy term [refer to the term proportional to $K_\perp$ in Eq.~(9) of Ref.~\cite{KISHINE20151}].

\bibliographystyle{apsrev4-1}
\bibliography{refs.bib}

\begin{thebibliography}{37}
\bibitem [{com()}]{comment:cont:B}%
  \BibitemOpen
  \href@noop {} {}\bibinfo {note} {We limit ourselves to studying the
  discontinuous $B$, sufficient to understand the qualitative behavior.
  Continuous $B$ can also be studied similarly, as shown in
  Ref.~\cite{KISHINE20151}.}\BibitemShut {Stop}%
\end{thebibliography}

\end{document}